\documentclass[12pt, letterpaper]{article}
\usepackage[top=1in, bottom=1.5in, left=1in, right=1in]{geometry}

\usepackage[utf8]{inputenc}
\usepackage{booktabs}
\usepackage{hyperref}
\usepackage{aas_macros}
\usepackage{color}
\usepackage{graphicx}
\usepackage{caption}
\usepackage{lineno}
\usepackage{deluxetable}
\usepackage{sidecap}

\title{Enhancing LSST Science with \emph{Euclid} Synergy}
\author{P. Capak, 
J-C. Cuillandre,
F. Bernardeau,
F. Castander,
R. Bowler,\\
C. Chang,
C. Grillmair, 
P. Gris,
T. Eifler,
C. Hirata,
I. Hook,
B. Jain,\\
K. Kuijken,
M. Lochner,
P. Oesch,
S. Paltani,
J. Rhodes, \\
B. Robertson,
D. Rubin,
R. Scaramella,
C. Scarlata,
D. Scolnic,\\
J. Silverman,
S. Wachter,
Y. Wang, \\
The Tri-Agency Working Group}

\date{November, 30, 2018}
\begin{document}

\maketitle

\begin{abstract}
This white paper is the result of the Tri-Agency Working Group (TAG) appointed to develop synergies between missions and is intended to clarify what LSST observations are needed in order to maximally enhance the combined science output of LSST and \emph{Euclid}.  To facilitate LSST planning we provide a range of possible LSST surveys with clear metrics based on the improvement in the Dark Energy figure of merit (FOM).  To provide a quantifiable metric we present five survey options using only between 0.3 and 3.8\% of the LSST 10 year survey.  We also provide information so that the LSST DDF cadence can possibly be matched to those of \emph{Euclid} in common deep fields, SXDS, COSMOS, CDFS, and a proposed new LSST deep field (near the Akari Deep Field South).

Co-coordination of observations from the Large Synoptic Survey Telescope (LSST) and \emph{Euclid} will lead to a significant number of synergies. The combination of optical multi-band imaging from LSST with high resolution optical and near-infrared photometry and spectroscopy from \emph{Euclid} will not only improve constraints on Dark Energy, but provide a wealth of science on the Milky Way, local group, local large scale structure, and even on first galaxies during the epoch of reionization.  A detailed paper has been published on the Dark Energy science case \cite{LSST-Euclid} by a joint LSST/\emph{Euclid} working group as well as a white paper describing LSST/\emph{Euclid}/WFIRST synergies \cite{jain}, and we will briefly describe other science cases here.  A companion white paper argues the general science case for an extension of the LSST footprint to the north at airmass $< 1.8$, and we support the white papers for southern extensions of the LSST survey. We also endorse the white papers from the LSST Dark Energy Science Collaboration (DESC) arguing for modifications to the Wide Fast Deep (WFD) and Deep Drilling Field (DDF) surveys, the Big Sky white paper, and the white paper from the WFIRST team arguing for a deep field on the Akari Deep Field South.

\end{abstract}

\clearpage
\section{White Paper Information}
Peter Capak, capak@caltech.edu

\begin{enumerate} 
\item {\bf Science Category:} The Nature of Dark Matter and Understanding Dark Energy, Exploring the Changing Sky, Milky Way Structure \& Formation.
\item {\bf Survey Type Category:} Mini Survey, Deep Drilling Field
\item {\bf Observing Strategy Category:}
A specific pointing or set of pointings that is (relatively) agnostic of the detailed observing strategy or cadence.
\end{enumerate}  

\clearpage

\section{Scientific Motivation}
Constraining the nature of Dark Matter and Dark Energy and improving measurements of the Milky Way are two of the key science goals of LSST.  With a relatively small (from 0.3 to 3.8\%) augmentation of the LSST survey to overlap the \emph{Euclid} survey, LSST will improve the constraints on Dark Energy by 30-69\% and improve the mapping of the milky-way halo by an order of magnitude over 50\% more area.  These gains come from combining the sensitive multi-color optical (0.35-1$\mu$m) imaging from LSST with the high-resolution optical and near-infrared (0.5-2$\mu$m) imaging and spectroscopy from \emph{Euclid}.  Here we outline a series of ranked surveys that will significantly improve the overall return of LSST science. \\

\begin{figure}
\includegraphics[width=16cm, height=6cm]{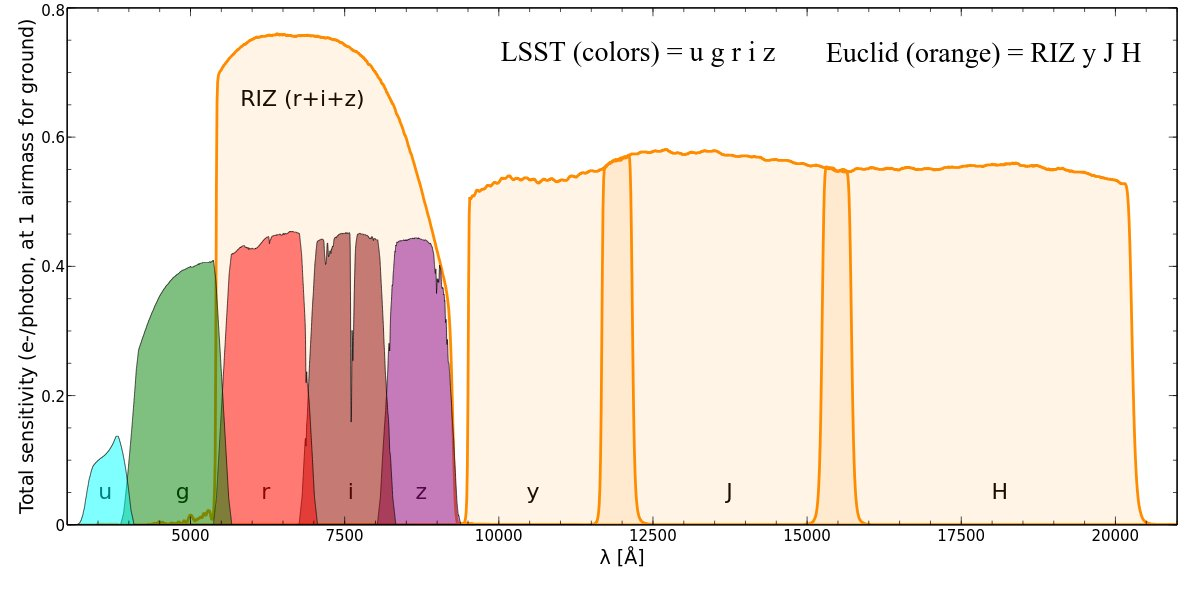}
\captionof{figure}{The planned imaging filters from LSST (orange) and \emph{Euclid} (blue) are shown.  \emph{Euclid} will provide 0.16$^{\prime\prime}$ FWHM imaging in a visible filter and similar quality 1-2$\mu$m near-infrared imaging that will complement LSST.  Maximizing the overlap between these two data sets will optimize the science return of both projects. }
\label{fig:filters}
\end{figure}

{\noindent \bf The \emph{Euclid} Survey:}
\emph{Euclid }is a 1.2m space-based telescope  scheduled for launch in mid 2022 with an expected mission life of 6.5 years.  It is designed to elucidation of the nature of Dark Energy (along with possible modifications to general relativity) using weak lensing and baryon acoustic oscillations (BAO) as probes.  To this end it will image $15,000$\
square degrees of sky in one broad optical band (VIS = $r+i+z$ covering $\sim$5400-9000 \AA)
with image quality (0.16$^{\prime\prime}$ FWHM) to an
expected depth of at least $24.5$ AB magnitudes (10$\sigma$ 1$^{\prime\prime}$~extended
source) and three near-infrared bands covering $\sim0.95-2 \mu$m down to
$Y,J,H=24.0$ AB magnitudes (5$\sigma$, point source) (Figure \ref{fig:filters}).
\emph{Euclid} will also obtain near-infrared low-resolution ($R\sim250$)
slitless spectroscopy using a ``red'' grism (1.25 to 1.85 $\mu$m).  

In addition to the main wide field survey \emph{Euclid} will image a series of deep and calibration fields to between 1.75-2 magnitudes deeper than the main survey in both imaging and the ``red” grism in addition to a ($R\sim250$) ``blue” grism (0.92 to 1.25 microns). The location of the deep fields were chosen to optimize synergy with the LSST survey whenever possible with the COSMOS, SXDS, CDFS, and the Akari Deep Field South or a field nearby chosen.\\

{\noindent \bf The Need for Survey Co-coordination:}
A cosmology survey that is optimized within a fixed time should cover the lowest galactic extinction parts of the sky prioritizing areas where the instrument has maximum sensitivity to detecting galaxies.  Since \emph{Euclid} is in space and LSST is on the ground there are different optimizations for maximum sensitivity.  In space, the entire sky is visible, and the primary background for \emph{Euclid} is the zodiacal light.  So sensitivity varies strongly with ecliptic latitude, exponentially decreasing as observations approach the ecliptic plane.  In contrast the primary factor for LSST sensitivity is the location on earth and atmosphere, which limits which areas of the sky can be observed and drives optimization to lower airmass with no strong dependence on ecliptic latitude. Despite these different optimization's Rhodes et al. \cite{LSST-Euclid} and Jain et al. \cite{jain} show there are several clear advantages to having LSST and \emph{Euclid} observe the same regions of the sky despite sub-optimal observing in some cases.  The primary benefit comes from maximizing the overlap of shallow LSST observations with the \emph{Euclid} wide survey.  The combination of these data will improve cosmology, studies of the milky-way halo and local group by improving statistics and mapping a larger part of the halo respectively.

The Dark Energy Science Collaboration (DESC) Wide Fast Deep (WFD) and the Big Sky white paper provides similar, but LSST specific, arguments for modifying the WFD survey to better match the extra-galactic sky.  A further extension north and south will provide even greater enhancements and a companion white paper gives a more general case for extending LSST coverage to $+2<$DEC$<+30$ beyond the \emph{Euclid} area.\\

\begin{SCfigure}
\centering
\caption{A comparison of spectroscopic and photometric redshifts for a representative set of 25,092 high-quality spectroscopic redshifts from the C3R2 survey \cite{c3r2} with real data comparable to that which will be obtained with LSST in Wide Fast Deep (WFD) and \emph{Euclid} wide surveys.  It is clear that combining \emph{Euclid} and LSST data significantly improves the photometric redshift performance.  We show standard photometric redshift performance metrics as well as those used by \cite{graham2017} for LSST. The optical imaging comes from CFHT-LS deep fields in u,g,r,i,z that are deeper than those that will be obtained by the LSST WFD survey.  The CFHT-LS photometry were degraded to the expected WFD depth in consultation with the LSST project.  The near-infrared imaging comes from the VISTA VIDEO and UltraVISTA surveys in Y,J,H,K bands that are comparable to what will be obtained by \emph{Euclid}. }  
\label{fig:pz-improve}
\includegraphics[height=11cm, width=0.5\columnwidth]{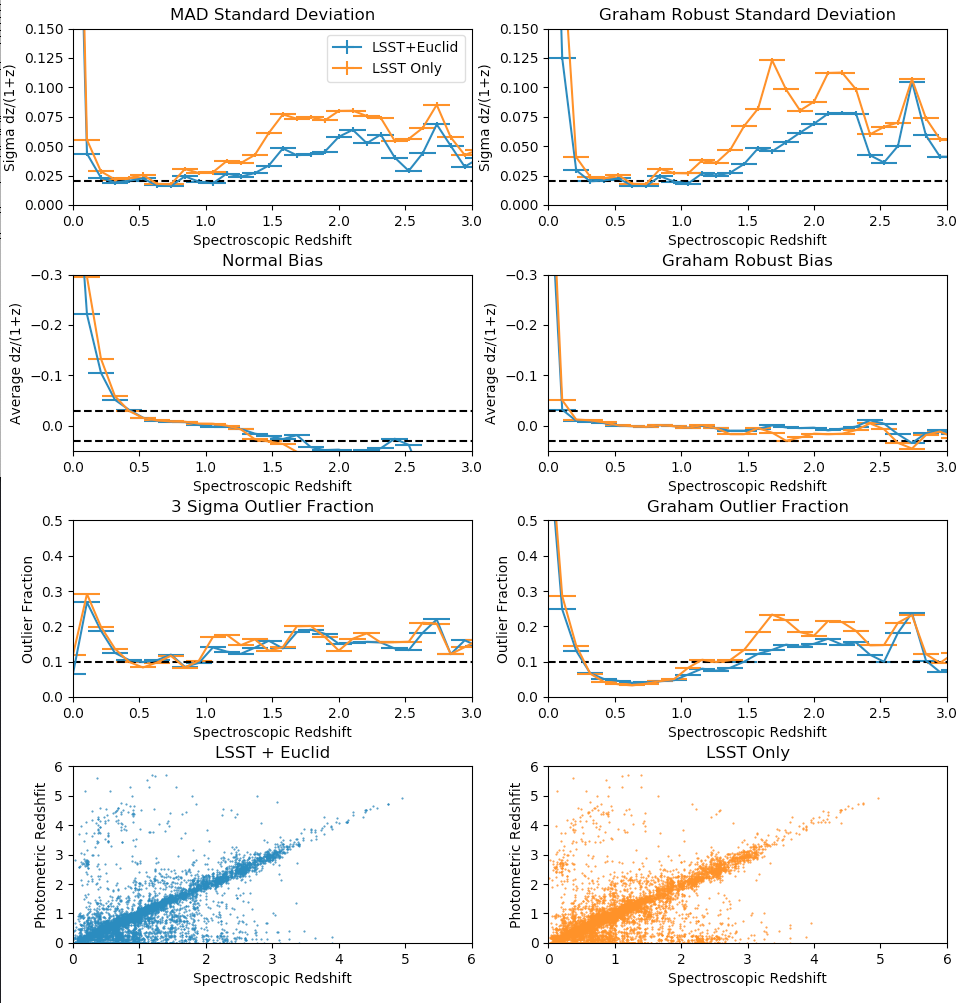}
\end{SCfigure}

{\noindent \bf Cosmology:}
A detailed science case for LSST-\emph{Euclid} cosmology is given in Rhodes et al. \cite{LSST-Euclid}.  In brief, the main gain to cosmology science comes from a combination of better photometric redshifts, improved weak lensing shear measurements, and improved de-blending of galaxy photometry.  Figure \ref{fig:pz-improve} shows an estimate of the improvement in photometric redshifts based on real data LSST like data. A further enhancement will come from observing a larger area that overlaps with the Dark Energy Spectroscopic Instrument (DESI) survey, and that can be observed by the Subaru Prime-Focus Spectrograph (PFS) which are both in the north.  The DESI data can be used to calibrate the photometric redshift measurements, conduct joint cosmology probes, and both DESI and PFS can carry out dedicated follow-up observations. 

Finally, by sharing common deep fields both LSST and \emph{Euclid} will benefit from spectroscopic and multi-wavelength imaging data taken to calibrate the surveys.  For this reason \emph{Euclid} chose to place calibration fields in COSMOS, SXDS, and VVDS, which are sub-optimal for \emph{Euclid} but enhance synergy with LSST. In addition, \emph{Euclid} has selected a $20$ square degrees deep field near the south ecliptic pole (near the Akari Deep Field South), which is well-optimized for \emph{Euclid} observations, and would greatly benefit from matching LSST observations for the same reasons.

We also note SNe Ia cosmology would greatly benefit from co-coordination in time. This would allow detection of live transients by both facilities quasi-simultaneously, providing information on optical-to-NIR color and location of the transient within its host galaxy. For Type Ia SNe this information has been shown to improves their use as standard candles. 
\\

{\noindent \bf Milky Way:}
The addition of high-resolution \emph{Euclid} imaging also greatly improves LSST star/galaxy separation to the \emph{Euclid} $5\sigma$ detection limit of $I\sim25.2 AB_{mag}$.  The improvement afforded by combining LSST and \emph{Euclid} data will improve sensitivity to low-surface brightness structures such as stellar streams and remnant or extended dwarf galaxies in the local group by at least an order of magnitude.  The proposed survey will increase the sky area where these studies can be carried out by 50\% from $\sim7000$ to $\sim10,500$ square degrees.  Furthermore, future DESI and PFS follow-up could be used to make dynamical maps.\\
 
{\noindent \bf Galaxy Evolution and Re-ionization:}
The addition of \emph{Euclid} near-infrared bands to LSST will extend the ability to measure stellar masses from $z\sim1.3$ to $z\sim3.5$, beyond the peak of the global star formation rate at $z\sim2-3$.  Furthermore, the addition of \emph{Euclid} spectra will provide line measurements for many LSST objects, especially in the Deep Drilling Fields (DDFs). The combination of \emph{Euclid} and LSST data will enable the discovery of luminous galaxies at $7<z<12$ that serve as signposts for the first ionized bubbles in the Universe. Based on current estimates \cite{bouwens,oesch,ono} we expect between $30-500$ galaxies and $\sim70$ AGN per 1000 square degrees and $\sim20,000$ galaxies in the deep fields \cite{bowler}. 

\clearpage

\clearpage
\section{Technical Description}

We propose two modifications to the LSST survey plan: {\bf 1)} Extend the LSST survey to cover the \emph{Euclid}/Space-Optimal low-extinction extra galactic sky to a depth that will enable statistics limited cosmology. {\bf 2)} Placing the last DDF in the \emph{Euclid} southern deep field near the Akari Deep Field South and covering a 20 square degree area around this field to moderate depth.\\ 
Assuming a cosmology survey is not systematics limited, the primary metric for this survey is overlapping \emph{Euclid} and LSST area. In particular, the Dark Energy figure of merit (FOM) increasing linearly with area.  A second improvement comes from improvements in the photometric redshifts, with the FOM increasing as the square root of the precision.  The baseline \emph{Euclid} photo-z precision requirements are a $\sigma_{pz}$ = dz/(1+z)$<$0.05 with simulations and data \cite{laigle} showing the precision increases with the signal-to-noise of the photometry linearly in the noise limited regime where the faint lensing galaxies reside. \\

In Table \ref{tab:surveys} we propose a range of surveys ranked by increase in FOM per visit that would enhance cosmology measurements from a baseline minimum survey to a maximal survey beyond which there are diminishing returns to cosmology. We note that full simulations including systematics would be required to derive an exact FOM and the numbers here are intended to provide a simple scaling for planning purposes only.  These constitute a ranked list of our proposed LSST survey modifications that can be used to evaluate synergy with other proposed surveys.\\

In Table \ref{tab:cosmo-improve} we provide a metric for improvement in the FOM scaling with the property listed.  The cost estimates in this table ares scaled form the DESC WFD survey simulation and reflect the cost compared with the Baseline2018a $-62<$DEC$<+2$ and expanded $-70<$DEC$<+12.5$ WFD survey.  We also note that the proposed Big Sky (K. Olsen white paper) strategy would be acceptable for Euclid.\\

Surveys 1 and 2 in Table \ref{tab:surveys} correspond to LSST surveys in the North and South that yield the minimum required photo-z precision for \emph{Euclid}. These provide the greatest gains by increasing the \emph{Euclid} area with LSST imaging. Survey 3-5 improve photo-z precision by adding u band and depth over the \emph{Euclid} area not in the WFD survey. \\

Finally, we propose placing the last DDF near the Akari Deep Field South to create significant scientific and calibration synergies.  Since the Euclid Deep Field is 20 square degrees in the south, we also propose a modest extended deep near or around the DDF to maximize synergy.  Furthermore, if possible the cadence of Euclid and LSST in all common deep fields (COSMOS, SXDS, CDFS, and the new Akari field) should be matched to further enhance transient science. 

\begin{deluxetable}{clcc}
\tabletypesize{\footnotesize}
\tablewidth{0pt}

 \tablecaption{Proposed Surveys \label{tab:surveys}}
  \tablehead{
 \colhead{Survey} & \colhead{Description} & \colhead{Area} & \colhead{N Visits} \\
  \colhead{Number} & \colhead{} & \colhead{Deg$^2$} & \colhead{per Point} 
 }
 \startdata 
1 & g+r+i+z in \emph{Euclid} Area at $+2<$DEC$<+30$ & 3,200 & 25 \\
2 & g+r+i+z in \emph{Euclid} Area at DEC$<-62$ & 1,200 & 25 \\
3 & u to non WFD \emph{Euclid} areas & 4,400 & 15 \\
4 & Increase g+r+i+z depth by 0.75 Mag & 4,400 & 83 \\
5 & u to WFD depth & 4,400 & 62 \\
Deep & u+g+r+i+z in \emph{Euclid}/Akari Deep Field & 20 & 1507\\ 
\enddata
\end{deluxetable}

\begin{deluxetable}{cccll}
\tabletypesize{\footnotesize}
\tablewidth{0pt}

 \tablecaption{ Improvements in Cosmology \label{tab:cosmo-improve}}
 \tablehead{
 \colhead{Survey} & \colhead{\% of 10yr} & \colhead{Improvement} & \colhead{Scales}& \colhead{Notes}\\
  \colhead{Number} &  \colhead{Survey} & \colhead{in FOM (Cumulative)} & \colhead{With} & \colhead{}
 }
 \startdata 
1 & 0.30-0.43\% & 1.30 (1.30) & Area/10,600 + 1 & Add 3,200 Deg$^2$ \\
2 & 0.09-0.16\% & 1.09 (1.42) & Area/13,800 + 1 & Add 1,200 Deg$^2$ \\
3 & 0.23-0.36\% & 1.03 (1.45) & $\sqrt{\sigma_{pz}}$Area/15,000 + 1 & 1.2\,$\times$ Photo-z improvement\\
4 & 0.91-1.37\% & 1.12 (1.63) & $\sqrt{\sigma_{pz}}$Area/15,000 + 1 & 2.0\,$\times$ Photo-z improvement\\
5 & 0.87-1.33\%  & 1.03 (1.69) & $\sqrt{\sigma_{pz}}$Area/15,000 + 1 & 1.2\,$\times$ Photo-z improvement\\
Deep & 0.08-0.16\% & Systematics & Area &  Calibration field\\
\enddata
\end{deluxetable}

\subsection{High-level description}

We propose shallow imaging in g,r,i,z and possibly u bands of the \emph{Euclid} wide footprint where LSST can observe at airmass $<1.8$.  Most of this area is covered by the WFD survey either in the Baseline2018a $-62<$DEC$<+2$ or the modified $-72<$DEC$<+12.5$ survey, and we are proposing a modest shallow extension of that survey.  \\

In addition, we are proposing DDF observations of the \emph{Euclid} Deep Field South located near the Akari Deep Field South.  This would consist of placing the last DDF at this location, and a shallower deep field covering in total 20 square degrees at this area.  Ideally this larger 20 square degree area would be a disk as currently planned by \emph{Euclid}, including the main deep-drilling field, but alternative shapes could also be considered in consultation with the \emph{Euclid} Survey planning team.\\

\subsection{Footprint -- pointings, regions and/or constraints}

The footprint of the wide survey is shown in Figure \ref{fig:footprint} and the \emph{Euclid} wide polygons are attached to this white paper as RA/DEC pairs formatted for easy import into python. \\

For planning purposes you can assume the \emph{Euclid} Deep field south pointings will be one LSST FOV at or around RA${^\circ}$=4:51:00 DEC${^\circ}$=-52:55:00 and RA${^\circ}$=4:35:00 DEC${^\circ}$=-54:40:00.  The exact position/geometry will need to evolve in the final survey planning in conjunction with LSST so these should not be taken as exact to more than a few degrees.  Furthermore, we note a single 20 deg. circle would be the preferred geometry for \emph{Euclid} but could be adapted in consultation with LSST.\\

\begin{minipage}{0.99\columnwidth}
\includegraphics[width=1.0\columnwidth]{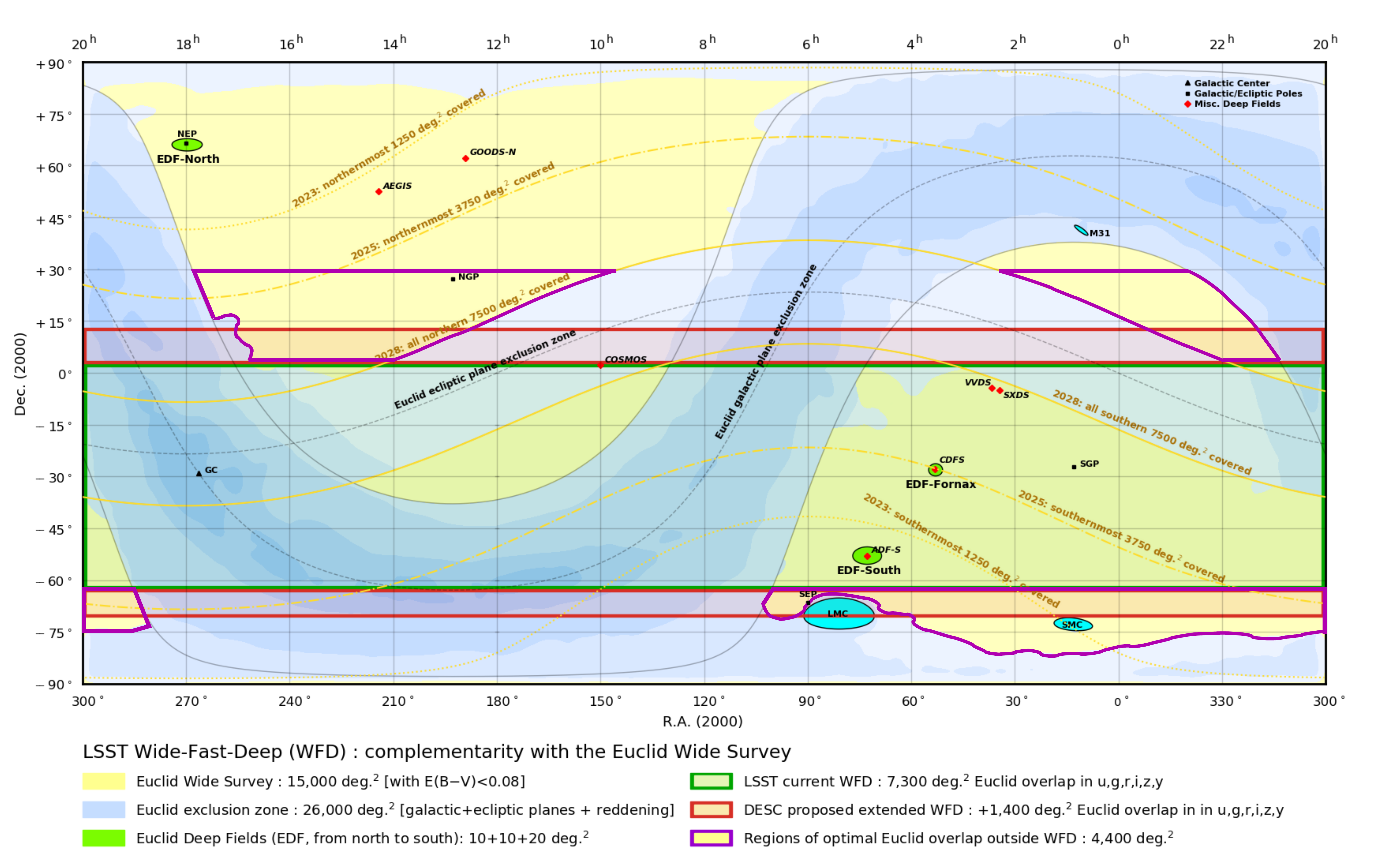}
\captionof{figure}{The proposed survey footprint is highlighted in purple and the notional Wide Fast Deep (WFD) survey is shaded in green.  The proposed extension to WFD from the DESC white paper is shaded in red.  The overall \emph{Euclid} footprint is marked in yellow with dashed lines indicating the progression of the survey with time (starting from the ecliptic poles).  The southern part of our proposed survey will be done in late 2022, but its northern part (in purple) will not start until 2025, and be complete in 2028.  The wide survey will be covered with the VIS (r+i+z) and Y,J,H filters as well as the 1.25-1.85$\mu$m $R\sim250$ grism.  The location of the \emph{Euclid} Deep Fields which will be 2 magnitudes deeper than the main survey are marked in bright green.  The location of calibration fields which will be 1.75 magnitudes deeper are marked with a red diamond.  The deep and calibration areas will be covered by a 0.9-1.25$\mu$m and a 1.25-1.85$\mu$m grism.   The calibration fields will be imaged to the full depth in the first year (2023).  The deep fields will accumulate depth over time following the cadence showing in Figure \ref{fig:cadence} from \cite{euclid-cadence}. }
\label{fig:footprint}
\end{minipage}

\subsection{Image quality}
The image quality should match the WFD survey image quality constraints in the r and i bands used for weak lensing to calibrate the shape and deblending measurements. This image quality would maximize the return for LSST.  However, our analysis indicates Pan-STARRS like seeing of 1-1.5$^{\prime\prime}$ is sufficient for \emph{Euclid} science alone where only color estimates are needed. 

\subsection{Individual image depth and/or sky brightness}

There are no constraints on this as long as the sensitivity estimates are met.

\subsection{Co-added image depth and/or total number of visits}

We request the depths listed in Table \ref{tab:depth} and a minimum of 5 exposures per point on the sky to ensure uniform depth and systematics control. \\

The depths in the deep fields are designed to differentiate between classes of very red objects detected at 26th magnitude in the \emph{Euclid} NIR bands. In particular, to exclude lower redshift contaminants from very high-redshift selections.  The primary requirement is a $5\sigma$ detection limit of 27 in the i and z bands, which is 1 magnitude deeper than the \emph{Euclid} NIR limit (2 magnitudes at 2$\sigma$).  The u,g,r and y depths are matched to the i,z and Euclid bands to provide optimal calibration of the WFD survey and synergy science.  

\begin{deluxetable}{ccccccccccccc}
\tabletypesize{\footnotesize}
\tablewidth{0pt}

 \tablecaption{ Depth Requirements, 5$\sigma$ point source \label{tab:depth}}
 \tablehead{
 \colhead{Survey} & \colhead{u} & \colhead{g} & \colhead{r} & \colhead{i} & \colhead{z} & \colhead{y}
 }
 \startdata 
1 & N/A & 25.7 & 25.1 & 24.8 & 24.6 & N/A \\
2 & N/A & 25.7 & 25.1 & 24.8 & 24.6 & N/A \\
3 & 25.4 & N/A & N/A & N/A & N/A & N/A \\
4 & N/A & 26.5 & 25.9 & 25.6 & 25.4 & N/A \\
5 & 26.2 & N/A & N/A & N/A & N/A & N/A \\
Deep & 27.5 & 27.5 & 27.0 & 27.0 & 27.0 & N/A \\
\enddata
\end{deluxetable}

\subsection{Number of visits within a night}
No constraints for the \emph{Euclid} wide field cosmology science case.  The deep field science case should be coordinated with the \emph{Euclid} observing cadence and have sufficient numbers of exposures to characterize transients.  Co-coordination with the wide field cadence may also yield additional transient science return.   We suggest doing this through simulation coordinated with DESC and the Euclid project.

\subsection{Distribution of visits over time}

Due to observing constraints the \emph{Euclid} survey will start at the ecliptic poles in late 2022 and move towards the ecliptic plane by 2029.  Based on the current best estimate observing plan this means it would be best to have the southern extension done as soon as possible, however this area is already covered by shallow DES data, so later coverage is OK, but sub-optimal.  \emph{Euclid} does not require the northern extension until it reaches that area of the sky, which will begin in late 2025 and does not need to be completed until 2028. However if the coverage occurs between 2025 and 2028 it should proceed from lowest to highest ecliptic latitude. \\

The deep field coverage should be co-coordinated to be coeval with the \emph{Euclid} cadence detailed in \cite{euclid-cadence} for the \emph{Euclid} Deep Field South (EDS, Akari Deep Field), and \emph{Euclid} Deep Field Fornax (EDF, CDFS) if possible to enhance transient science.  In addition to these two deep fields the SXDS and COSMOS fields will be observed with high-cadence in the early part of the \emph{Euclid} mission.  If possible the cadence should be co-coordinated.  However, the exact timing of these observations will depend on the launch date, so detailed plans should be co-coordinated with \emph{Euclid}. \\

We also note distributing the requested data in g and r band over a long time baseline will enable proper-motion measurements which would be valuable for studies of galactic streams and dwarfs as well as transient phenomenon.\\ 

\begin{minipage}{0.99\columnwidth}
\includegraphics[width=1.0\columnwidth]{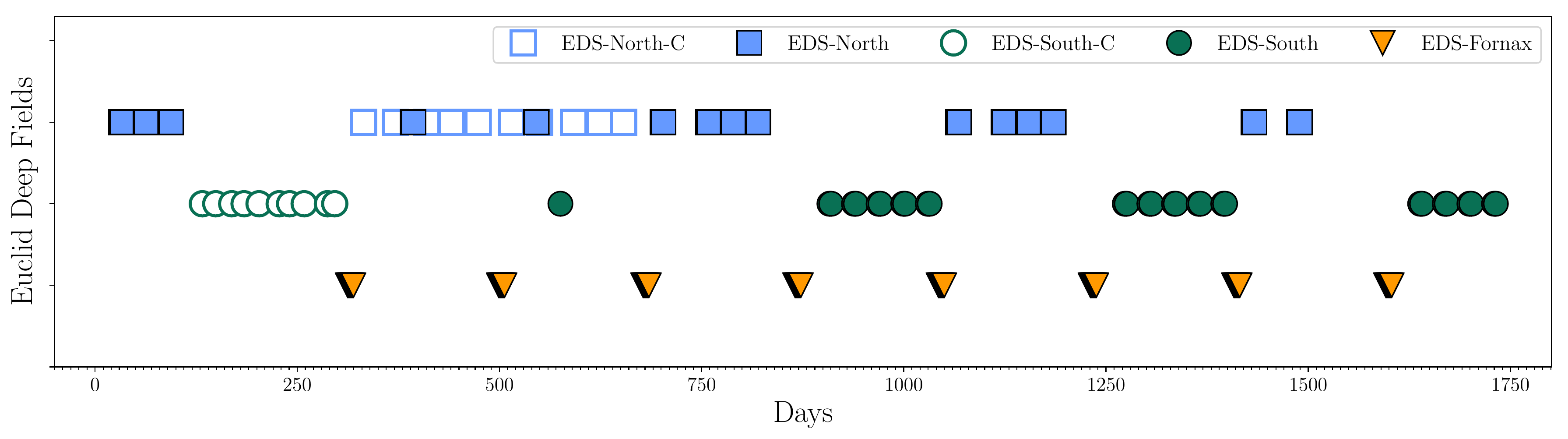}
\captionof{figure}{This is Figure 1 from \cite{euclid-cadence} showing the cadence of the \emph{Euclid} deep fields, the North Ecliptic Pole (EDS-North and EDS-North-C), the Akari Deep Field south (EDS-South and EDS-South-C) and the Chandra Deep Field South (EDS-Fornax).  The -C designation denotes the grism de-contamination observations which must be distributed over a range of position angels.   The exact cadence will depend on the launch date of \emph{Euclid}. To optimize transient science the LSST and \emph{Euclid} Cadence should be optimized as much as possible.  }
\label{fig:cadence}
\end{minipage}

\subsection{Filter choice}
In the wide fields, our priority is g,r,i,z coverage.  u band coverage would improve our photometric redshifts, so it has next priority. We do not require y band coverage although it would enhance other science cases.  \\

In the deep fields we require u band to provide calibration to the overall WFD survey.  However, we still exclude y band because it is largely redundant with the Euclid Y band.  While y band would provide improved high-redshift galaxy selection at $z\sim7-8$ where the \emph{Euclid} and LSST bands overlap, enabling a finer constraint in redshift, obtaining a useful depth would require 3,050 exposures which is nearly twice as much time as all the other bands combined. 

\subsection{Exposure constraints}

We have no constraints on the number of exposures per visit as long as a grand total of at least 5 exposures is reached to get to the required depths. The 5 exposure depth is set to achieve even coverage and remove position dependent systematics.  We prefer the standard two exposures per visit to remove cosmic rays but this is not required.

\subsection{Other constraints}
None.

\subsection{Estimated time requirement}
We estimate the time based on the DESC WFD white paper simulations scaling to the number of total visits and area of each survey.  We assume the WFD survey has 825 visits per pointing, covers 18,000 square degrees, and uses 80\% of the total survey.  We also assume a minimum of 5 visits per filter to control relative photometric calibration systematics.\\

We assume two cases, one where the Baseline2018a $-62<$DEC$<+2$ WFD survey is adopted and the remaining DDF is not placed on the \emph{Euclid} Deep field and a second option where the proposed $-70<$DEC$<+2$ WFD is adopted and the remaining DDF is placed on the \emph{Euclid} Deep Field south (Akari Deep Field). \\

Table \ref{tab:time} gives the estimated survey time for each of the proposed surveys under the two assumptions.  The number of visits per band in this table is differential, assuming the previous survey has been completed so the new survey is just adding time/visits.  In other words, the surveys should be summed to get total visits required.  For clarity the cumulative percentages are given for each survey in brackets. 

\begin{deluxetable}{cccccc}
\tabletypesize{\footnotesize}
\tablewidth{0pt}

 \tablecaption{Time Estimates for Each Survey\label{tab:time}}
 \tablehead{
 \colhead{Survey} & \colhead{Visits} & \colhead{Area wrt Current} &  \colhead{Fractional} & \colhead{Area wrt Proposed} &  \colhead{Fractional Time}\\
  \colhead{} & \colhead{(u,g,r,i,z)} & \colhead{WFD (Deg$^2$)} &  \colhead{Time (\%)} & \colhead{WFD (Deg$^2$)} &  \colhead{new WFD (\%)}
 }
 \startdata 
1 & 25 (0,5,5,5,10) & 3,200 & 0.43 (0.43) & 2,200 & 0.30 (0.30)\\
2 & 25 (0,5,5,5,10) & 1,200 & 0.16 (0.59)& 700 & 0.09 (0.39)\\
3 & 15 (15,0,0,0,0) & 4,400 & 0.36 (0.95) & 2,900 & 0.23 (0.62)\\
4 & 58 (0,10,5,13,30) & 4,400 & 1.37 (2.23) & 2,900 & 0.91 (1.53)\\
5 & 56 (56,0,0,0,0) & 4,400 & 1.33 (3.65)&  2,900 & 0.87 (2.41)\\
Deep & 1507 (738,96,73,266,334) & 20 & 0.16 (3.81) & 10 & 0.08 (2.49)\\
\enddata
\end{deluxetable}

\begin{table}[ht]
    \centering
    \begin{tabular}{l|l|l|l}
        \toprule
        Properties & Importance \hspace{.3in} \\
        \midrule
        Image quality & 2    \\
        Sky brightness &  3 \\
        Individual image depth &  3 \\
        Co-added image depth &   1 \\
        Number of exposures in a visit   &  3 \\
        Number of visits (in a night)  &  3 \\ 
        Total number of visits &   2 \\
        Time between visits (in a night) & 3 \\
        Time between visits (between nights)  &  3 \\
        Long-term gaps between visits & 3 \\
        Other (please add other constraints as needed) & 3 \\
        \bottomrule
    \end{tabular}
    \caption{{\bf Constraint Rankings:} Summary of the relative importance of various survey strategy constraints. Please rank the importance of each of these considerations, from 1=very important, 2=somewhat important, 3=not important. If a given constraint depends on other parameters in the table, but these other parameters are not important in themselves, please only mark the final constraint as important. For example, individual image depth depends on image quality, sky brightness, and number of exposures in a visit; if your science depends on the individual image depth but not directly on the other parameters, individual image depth would be `1' and the other parameters could be marked as `3', giving us the most flexibility when determining the composition of a visit, for example.}
        \label{tab:obs_constraints}
\end{table}

\subsection{Technical trades}

{\it What is the effect of a trade-off between your requested survey footprint (area) and requested co-added depth or number of visits?}
\\

Our primary requirement is the footprint, followed by depth, followed by number of visits. 
\\
\\
    
{\it If not requesting a specific timing of visits, what is the effect of a trade-off between the uniformity of observations and the frequency of observations in time? e.g. a `rolling cadence' increases the frequency of visits during a short time period at the cost of fewer visits the rest of the time, making the overall sampling less uniform.}
\\

Our primary requirement is uniformity which should reach the requested depth over the whole area by 2028.  For transient science it would be valuable if the timing were matched in the deep fields.\\
\\
{\it What is the effect of a trade-off on the exposure time and number of visits (e.g. increasing the individual image depth but decreasing the overall number of visits)?}
\\

Depth is the primary consideration, the number of exposures improves artifact rejection and systematics and could be as low as 5 per band. 
\\
\\
{\it What is the effect of a trade-off between uniformity in number of visits and co-added depth? Is there any benefit to real-time exposure time optimization to obtain nearly constant single-visit limiting depth?}
\\

The total depth is the required metric, there is little benefit to single exposure matched depth. 
\\
\\
{\it Are there any other potential trade-offs to consider when attempting to balance this proposal with others which may have similar but slightly different requests?}
\\

Our primary consideration is the requested footprint, followed by the requested depth, followed by the request to complete by 2028.
\\

\section{Performance Evaluation}
 Our figure of merit is given in Table \ref{tab:surveys} based on improvements to the Dark Energy FOM from joint LSST/Euclid data.  This scales as the area covered until the full LSST accessible area is covered.  It then scales as the square root of the depth of coverate with respect to the surveys presented in Table \ref{tab:cosmo-improve} and depths presented in Table \ref{tab:depth}.


\section{Special Data Processing}

None.

\pagebreak

\section{References}
\bibliographystyle{hunsrt}
\begingroup
\renewcommand{\section}[2]{}%
\bibliography{ms}

\begin{thebibliography}{10}

\bibitem{LSST-Euclid}
J.~{Rhodes}, R.~C. {Nichol}, {\'E}.~{Aubourg}, et~al.
\newblock {Scientific Synergy between LSST and Euclid}.
\newblock {\em \apjs}, 233:21, 2017, \href{http://arxiv.org/abs/1710.08489}{\tt
  arXiv:1710.08489}.

\bibitem{jain}
B.~{Jain}, D.~{Spergel}, R.~{Bean}, et~al.
\newblock {The Whole is Greater than the Sum of the Parts: Optimizing the Joint
  Science Return from LSST, Euclid and WFIRST}.
\newblock {\em ArXiv e-prints}, page arXiv:1501.07897, 2015,
  \href{http://arxiv.org/abs/1501.07897}{\tt arXiv:1501.07897}.

\bibitem{bouwens}
R.~J. {Bouwens}, G.~D. {Illingworth}, M.~{Franx}, and H.~{Ford}.
\newblock {UV Luminosity Functions at z\~{}4, 5, and 6 from the Hubble Ultra
  Deep Field and Other Deep Hubble Space Telescope ACS Fields: Evolution and
  Star Formation History}.
\newblock {\em \apj}, 670:928--958, 2007,
  \href{http://arxiv.org/abs/0707.2080}{\tt arXiv:0707.2080}.

\bibitem{oesch}
P.~A. {Oesch}, M.~{Montes}, N.~{Reddy}, et~al.
\newblock {HDUV: The Hubble Deep UV Legacy Survey}.
\newblock {\em \apjs}, 237:12, 2018, \href{http://arxiv.org/abs/1806.01853}{\tt
  arXiv:1806.01853}.

\bibitem{ono}
Y.~{Ono}, M.~{Ouchi}, Y.~{Harikane}, et~al.
\newblock {Great Optically Luminous Dropout Research Using Subaru HSC
  (GOLDRUSH). I. UV luminosity functions at z {$\sim$} 4-7 derived with the
  half-million dropouts on the 100 deg$^{2}$ sky}.
\newblock {\em \pasj}, 70:S10, 2018, \href{http://arxiv.org/abs/1704.06004}{\tt
  arXiv:1704.06004}.

\bibitem{bowler}
R.~A.~A. Bowler, J.~S. Dunlop, R.~J. McLure, and D.~J. McLeod.
\newblock {Unveiling the nature of bright z {\~{}} 7 galaxies with the Hubble
  Space Telescope}.
\newblock {\em MNRAS}, 466:3612--3635, 2017.

\bibitem{c3r2}
D.~C. {Masters}, D.~K. {Stern}, J.~G. {Cohen}, P.~L. {Capak}, J.~D. {Rhodes},
  F.~J. {Castander}, and S.~{Paltani}.
\newblock {The Complete Calibration of the Color-Redshift Relation (C3R2)
  Survey: Survey Overview and Data Release 1}.
\newblock {\em \apj}, 841:111, 2017, \href{http://arxiv.org/abs/1704.06665}{\tt
  arXiv:1704.06665}.

\bibitem{graham2017}
M.~L. {Graham}, A.~J. {Connolly}, {\v Z}.~{Ivezi{\'c}}, S.~J. {Schmidt}, R.~L.
  {Jones}, M.~{Juri{\'c}}, S.~F. {Daniel}, and P.~{Yoachim}.
\newblock {Photometric Redshifts with the LSST: Evaluating Survey Observing
  Strategies}.
\newblock {\em \aj}, 155:1, 2018, \href{http://arxiv.org/abs/1706.09507}{\tt
  arXiv:1706.09507}.

\bibitem{laigle}
C.~{Laigle}, H.~J. {McCracken}, O.~{Ilbert}, et~al.
\newblock {The COSMOS2015 Catalog: Exploring the 1 \&lt; z \&lt; 6 Universe
  with Half a Million Galaxies}.
\newblock {\em The Astrophysical Journal Supplement Series}, 224:24, 2016.

\bibitem{euclid-cadence}
C.~{Inserra}, R.~C. {Nichol}, D.~{Scovacricchi}, et~al.
\newblock {Euclid: Superluminous supernovae in the Deep Survey}.
\newblock {\em \aap}, 609:A83, 2018, \href{http://arxiv.org/abs/1710.09585}{\tt
  arXiv:1710.09585}.

\end{thebibliography}
\endgroup
\end{document}